\documentclass[pra,aps,twocolumn]{revtex4}
\usepackage{amsmath,amsfonts,amsthm,amssymb}
\usepackage{graphicx}
\usepackage{epstopdf}
\newcommand\C{{\mathbb C}}

\newcommand\Tr{{\rm Tr\, }}
\newcommand\half{{\mbox{$\frac 12$}}}
\newcommand\eps{\varepsilon}
\newcommand\const{{\rm const. \, }}

\newcommand{\Aa}{{\mathcal A}}
\newcommand{\Bb}{{\mathcal B}}
\newcommand{\beq}{\begin{equation}}
\newcommand{\eeq}{\end{equation}}
\newcommand{\p}{{\mathcal P}}
\newcommand{\vecp}{{\bf p}}
\newcommand{\vecq}{{\bf q}}
\newcommand{\0}{{\bf 0}}
\newcommand{\vecx}{{\bf x}}
\newcommand{\vecy}{{\bf y}}

\newtheorem{thm}{THEOREM}
\newtheorem{lem}{Lemma}

\begin{document}

\title{ Bose-Einstein Quantum Phase Transition in an
Optical Lattice Model}

\author{Michael Aizenman}
%\email{aizenman@princeton.edu}
\affiliation{Department of Physics, Jadwin Hall, Princeton University,
P.O. Box 708, Princeton NJ 08544, USA}

\author{Elliott H. Lieb}
%\email{lieb@princeton.edu}
\affiliation{Department of Physics, Jadwin Hall, Princeton University,
P.O. Box 708, Princeton NJ 08544, USA}

\author{Robert Seiringer}
%\email{rseiring@princeton.edu}
\affiliation{Department of Physics, Jadwin Hall, Princeton University,
P.O. Box 708, Princeton NJ 08544, USA}

\author{Jan Philip Solovej}
%\email{solovej@math.ku.dk}
\altaffiliation[On leave from ]{Department of Mathematics,
University of Copenhagen, Universitetsparken 5, DK-2100 Copenhagen,
Denmark}
\affiliation{School of Mathematics, Institute for Advanced Study, 1
Einstein Drive, Princeton NJ 08540, USA}

\author{Jakob Yngvason}
%\email{yngvason@thor.thp.univie.ac.at}
\affiliation{Institut f\"ur Theoretische Physik, Universit\"at Wien,
Boltzmanngasse 5, A-1090 Vienna, Austria}
\affiliation{Erwin Schr\"odinger Institute for Mathematical Physics,
    Boltzmanngasse 9, A-1090 Vienna, Austria}

\date{March 8, 2004}

\begin{abstract}
   Bose-Einstein condensation (BEC) in cold gases can be turned on and
   off by an external potential, such as that presented by an optical
   lattice. We present a model of this phenomenon which we are able to
   analyze rigorously. The system is a hard core lattice gas at
   half-filling and the optical lattice is modeled by a periodic
   potential of strength $\lambda$. For small $\lambda$ and
   temperature, BEC is proved to occur, while at large $\lambda$ or
   temperature there is no BEC. At large $\lambda$ the low-temperature
   states are in a Mott insulator phase with a characteristic gap that
   is absent in the BEC phase. The interparticle interaction is
   essential for this transition, which occurs even in the ground
   state.  Surprisingly, the condensation is always into the $\vecp=\0$
   mode in this model, although the density itself has the periodicity
   of the imposed potential.
\end{abstract}

%\pacs{}

\maketitle

\section{Introduction}

One of the most remarkable recent developments in the study of
ultracold Bose gases is the observation of a reversible transition
from a Bose-Einstein condensate to a state composed of localized atoms
as the strength of a periodic, optical trapping potential is varied
\cite{G1,G2}.  This is an example of a quantum phase transition
\cite{Sa} where quantum fluctuations and correlations rather than
energy-entropy competition is the driving force and its theoretical
understanding is quite challenging.  The model usually considered for
describing this phenomenon is the Bose-Hubbard model and the
transition is interpreted as the superfluid-insulator transition that
was studied in \cite{FWGF} with an application to ${\rm He}^4$ in
porous media in mind.  The possibility of applying this scheme to
gases of alkali atoms in optical traps was first realized in
\cite{JBCGZ}.  The article \cite{Z} reviews these developments and
many recent papers, e.g., \cite{GCZKSD,Ziegler,NS,G,Zi,DODS,RBREWC,MA}
are devoted to this topic.  These papers contain also further
references to earlier work along these lines.

The investigations of the phase transition in the Bose-Hubbard model
are mostly based on variational or numerical methods and the signal of
the phase transition is usually taken to be that an ansatz with a
sharp particle number at each lattice site leads to a lower energy
than a delocalized Bogoliubov state.  On the other hand, there exists
no rigorous proof, so far, that the true ground state of the model has
off-diagonal long range order at one end of the parameter regime that
disappears at the other end.  In the present paper we study a slightly
different model where just this phenomenon can be rigorously proved
and which, at the same time, captures the salient features of the
experimental situation.

The model is that of a hard core gas on a cubic lattice at
half-filling (i.e., when the particle number is half the number of
sites).  The `optical lattice' is modeled simply by a periodic,
one-body potential $\lambda (-1)^{\vecx}$, where $(-1)^{\vecx}=+1$ on the
A-sublattice and $(-1)^{\vecx}=-1$ on the B-sublattice.  Thus, the
Hamiltonian, expressed through bosonic creation and annihilation
operators, equals
\begin{eqnarray}\nonumber
H&=& - \half \sum_{\langle \vecx\vecy\rangle} ( a^\dagger_\vecx
a^{\phantom\dagger}_\vecy
+ a^{\phantom\dagger}_\vecx a^\dagger_\vecy ) +
\lambda \sum_\vecx (-1)^\vecx a^\dagger_\vecx a^{\phantom\dagger}_\vecx \\ &&
+ U \sum_\vecx a^\dagger_\vecx a^{\phantom\dagger}_\vecx
(a^\dagger_\vecx a^{\phantom\dagger}_\vecx-1). \label{hamiltonian}
\end{eqnarray}
The sites $\vecx$ are in a $d$-dimensional hypercubic lattice, and
$\langle \vecx\vecy\rangle$ stands for pairs of nearest neighbors. The case
considered in this paper is the hard-core interaction $U=\infty$.
Apart from the periodic potential, this is also the Bose-Hubbard model
with infinite on-site repulsion.

Note that in our model the troughs of the optical potential correspond
to the B-sublattice where the periodic potential is negative, and the
crests correspond to the A-sublattice.  Often in the Bose-Hubbard
model the whole lattice itself is used to approximate the troughs
alone.  Roughly speaking, half-filling in our model corresponds to
filling factor 1 in the Bose-Hubbard approximation.

As is well known, the model (\ref{hamiltonian}) with $U=\infty$ can
also be viewed as the XY model of a spin 1/2 system \cite{MM}. The
periodic potential then corresponds to a staggered magnetic field.
This will be explained in the next section.

We are able to prove the following facts rigorously for 3 or more
dimensions. The $\lambda$-$T$-phase diagram at half-filling (e.g.,
mean density $\varrho=\half$) is shown schematically in
Figure~\ref{fig1}.

\begin{enumerate}
\item If both $\lambda$ and the temperature $T$ are small, then there
    is Bose-Einstein condensation (BEC).  In this parameter regime the
    one-body density matrix has exactly one large eigenvalue (in the
    thermodynamic limit), and the corresponding condensate wave function
    is $\varphi(\vecx)= \const$
\label{bec}

\item If either $T$ or $\lambda$ is big enough, then the correlation
    function (the one-body density matrix) decays exponentially, and
    hence there is {\it no BEC}.  In particular, this applies to the
    ground state ($T=0$) for $\lambda$ big enough.
\label{nobec}

\item The Mott insulator phase is characterized by a gap, i.e., a jump
    in the chemical potential (at zero temperature). We are able to
    prove this, at half-filling, for big enough $\lambda$.  More
    precisely, there is a cusp in the dependence of the ground state
    energy on the number of particles; adding or removing one particle
    costs a non-zero amount of energy.  We also show that there is no
    such gap whenever there is BEC.\label{item5}

\begin{figure}[htf]
\includegraphics[width=7.5cm, height=5.5cm]{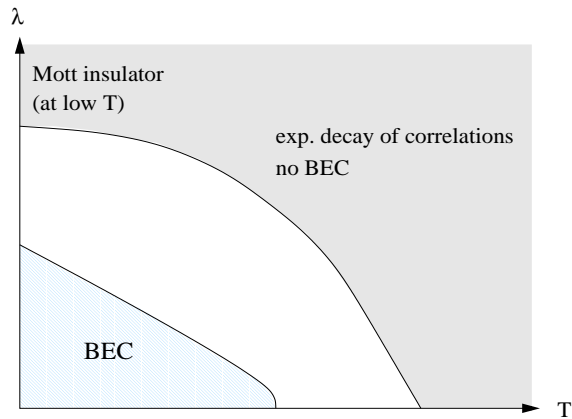}
\caption{Schematic phase diagram at half-filling}
\label{fig1}
\end{figure}

\item The interparticle interaction is essential for items~\ref{nobec}
    and~\ref{item5}.  Non-interacting bosons {\it always display BEC}
    for low, but positive $T$ (which depends on $\lambda$, of course).
\label{freegas}

\item For all $T\geq 0$ and all $\lambda > 0$ the diagonal part of the
    one-body density matrix $\langle a^\dagger_\vecx a^{\phantom
      \dagger}_\vecx \rangle$ is {\it not constant}. Its value on the
    A-sublattice is constant, but strictly less than its constant value
    on the B-sublattice (for a finite system with periodic boundary
    conditions) and this discrepancy survives in the thermodynamic
    limit. In contrast, in the regime mentioned in item~\ref{bec}, the
    off-diagonal long-range order is constant, i.e., $\langle
    a^\dagger_\vecx a^{\phantom \dagger}_\vecy \rangle \approx \const $
    for large $|\vecx-\vecy|$.

\end{enumerate}

We give explicit expressions for the curves, sketched in
Figure~\ref{fig1}, defining the regimes for which the above statements
are proved here (see Eqs. (\ref{beccurve}) and (\ref{eq:highLT})
below.)  We are not able to make a rigorous statement about the
intermediate regime, but we believe that there is only a critical line
separating the BEC and the Mott insulator phases.

We focus here on lattice dimensions $d\geq 3$ but, using the technique
employed in \cite{KLS}, an extension to the ground state in two
dimensions is possible. Other possible extensions are mentioned in the
next section.

\section{Detailed Description of the Model}\label{sectdet}

We write the Hamiltonian (\ref{hamiltonian}) of the lattice gas with $U=\infty$
in terms of the creation and annihilation operators, $a_{\vecx}^\dagger$
and $a^{\phantom\dagger}_{\vecx}$, for particles at lattice site $\vecx\in
\Lambda$, with $\Lambda$ a finite hypercubic lattice with $L^d$ sites,
$L$ being an even integer. We impose periodic boundary conditions.
Because of the hard-core condition, there is at most one particle at
each site, and thus the creation and annihilation operators can be
represented as $2\times 2$ matrices with
$$
a^\dagger_{\vecx}\leftrightarrow \left(
     \begin{array}{cc} 0 & 1 \\ 0 & 0
\end{array} \right) , \quad a^{\phantom\dagger}_{\vecx}\leftrightarrow
\left( \begin{array}{cc} 0 & 0 \\ 1 & 0 \end{array} \right) , \quad
a^\dagger_{\vecx}a^{\phantom\dagger}_\vecx\leftrightarrow \left(
\begin{array}{cc} 1 & 0 \\ 0 & 0 \end{array} \right),
$$
for each $\vecx\in\Lambda$. The correspondence with the spin
matrices
$$
S^1=\frac12 \left(
\begin{array}{cc} 0 & 1 \\ 1 & 0 \end{array} \right), S^2=\frac12\left(
\begin{array}{cc} 0 & -{\rm i} \\ {\rm i} & 0 \end{array} \right),
S^3=\frac12\left( \begin{array}{cc} 1 & 0 \\ 0 & -1 \end{array} \right)
$$
is
$$
a^\dagger_{\vecx}=S^1_{\vecx}+{\rm i}S^2_{\vecx}\equiv S_\vecx^+, \quad
a^{\phantom\dagger}_{\vecx}=S^1_{\vecx}-{\rm i}S^2_{\vecx}\equiv S_\vecx^-,
$$
and hence $a^\dagger_{\vecx}a^{\phantom\dagger}_{\vecx}=S_\vecx^3+\half$.
Adding a convenient constant to make the periodic potential positive,
the Hamiltonian (\ref{hamiltonian}) for $U=\infty$ is thus equivalent
to
\begin{eqnarray}\label{ham}
H&=&-\half \sum_{\langle \vecx\vecy\rangle}
(S^+_\vecx S^-_\vecy+S^-_\vecx S^+_\vecy)+\lambda\sum_\vecx \left[\half +
(-1)^\vecx S^3_\vecx\right]
\nonumber\\
&=&-\sum_{\langle \vecx\vecy\rangle}(S^1_\vecx S^1_\vecy+S^2_\vecx S^2_\vecy)
+\lambda\sum_\vecx\left[ \half + (-1)^\vecx S^3_\vecx\right].
\end{eqnarray}
As explained in the introduction, $(-1)^\vecx= \pm 1$ on alternating
sites.  Without loss of generality we may assume $\lambda\geq 0$. Note
that the subtraction of the \lq diagonal\rq\ terms in the kinetic
energy has the effect of a chemical potential and as a consequence the
unique ground state of (\ref{ham}) has particle number
$N=\half|\Lambda|$. We postpone the proof of this assertion to 
Appendix~\ref{apphalf}.

The presence or absence of Bose-Einstein condensation is expressed through
the reduced one-particle density matrix $$
\gamma(\vecx,\vecy)=\langle
a^\dagger_{\vecx}a^{\phantom\dagger}_{\vecy}\rangle
$$
where $\langle\cdot\rangle$ denotes the expectation value in the
thermal equilibrium state or the ground state considered. BEC occurs
(by definition) if the $L^d\times L^d$ matrix $\gamma(\vecx,\vecy)$ has an
eigenvalue of order $N$ in the thermodynamic limit $\Lambda\to\infty$,
$N\to\infty$, with $\varrho=N/|\Lambda|$ fixed.

We shall prove that for $\varrho=1/2$ and $d\geq 3$ the thermal
equilibrium state of \eqref{ham} shows Bose-Einstein condensation for
small $\lambda$ and low temperature $T$, while for large $\lambda$ or $T$ the
condensation disappears.  For $d=2$ this is true only in the ground
state.  Here, $\varrho$ stands for the average density, since we are using
the grand-canonical ensemble where the particle number is not fixed.
Note that we are not dealing here with a dilute system and the
condensation is always depleted, even in the ground state.

We remark that the Hamiltonian (\ref{ham}) is invariant under the following two
unitary transformations, which will be used throughout the paper.
    {\em 1)\/} Uniform rotation around the $S^3$-axis, in particular the
map $S_\vecx^{1}\to - S_\vecx^{1}$ and $S_\vecx^2\to -S^2_\vecx$ at
all sites. The
corresponding conserved generator is the total particle number.
{\em 2)\/}  The particle-hole symmetry, which corresponds to:
$S_\vecx^1\leftrightarrow S_\vecx^2$, $S^3_\vecx\to -S^3_\vecx$
at all sites, followed by a unit-vector translation in any of the
lattice directions.  For the latter symmetry the half-filling is
essential.

Our analysis of the system proceeds via the following steps. In
the next Section~\ref{becproof} we prove infrared bounds on the
two-point function
$\langle \widetilde S^1_\vecp \widetilde S^1_{-\vecp}+ \widetilde
S^2_\vecp\widetilde
S^2_{-\vecp}\rangle$ in momentum-space.  The essential ingredient
here is {\it reflection positivity} of the Hamiltonian \eqref{ham} and
the closely related property of {\it Gaussian domination}
\cite{DLS}. The bound obtained depends on $\lambda$ and the
temperature. A sum rule fixes the sum over $\vecp$ of the two-point
function and the infrared bound leads to the conclusion that for small
$\lambda$ and low temperatures the contribution from $\vecp=\0$ remains
non-vanishing in the thermodynamic limit. This proves BEC.

Reflection positivity is essential for our proof of BEC and forces the
periodic potential to have period 2.  Other generalizations can be
accommodated, however, such as a more general lattice in which we add
hopping to next nearest neighbors \cite{LS} or the addition of nearest
neighbor interparticle repulsion \cite{DLS}. For simplicity we
concentrate here on the simple cubic lattice with on-site repulsion
only.

In Section~\ref{sectgap} we show that the existence of BEC implies the
absence of an energy gap for adding or removing a particle, and that
the energy viewed as a function of the density has a unique tangent at
$\varrho=\half$, i.e., the chemical potential is continuous.

The absence of BEC for large $\lambda$ or high temperatures $T$ is
proved in Section~\ref{loops}. The technique applied here is a path
space representation of the two-point function that follows from the
Trotter product formula. This representation allows us to derive
exponential decay of the two-point function, provided $\lambda$ or $T$
are sufficiently large, and hence absence of long range order. The
magnitude of $\lambda$ or $T$ enters through the suppression in
the path space integral of long contours connecting a pair of lattice
points if these parameters are large. The same representation is used
for proving the existence of a gap, as explained in item~\ref{item5}
in the introduction. This method is quite robust and easily extends to
a periodicity of the optical lattice potential different from 2, for
instance.

In Section~\ref{stagdens} we show that the particle density (or
$3$-component of the spin) oscillates with the period of the staggered
field if $\lambda\neq 0$, in contrast to the condensate wave function
which is independent of $\lambda$ and $\vecx$. For this effect the
interaction is essential, as remarked in Section~\ref{sectfree}.
Without the interaction there is {\it always} BEC (for low $T$) and the condensate
wave function is {\it never} constant (for $\lambda\neq 0$).

In Appendix~\ref{apphalf}, we shall prove that the ground state of \eqref{ham}
has total spin 0, which in the lattice gas language means that the
lowest energy of \eqref{ham} is obtained when the particle number is
$|\Lambda|/2$. The essential ingredient in our proof is again
reflection positivity of the Hamiltonian \eqref{ham}. We also show
that the canonical partition function is maximal at half-filling.

\section{Proof of BEC for small $\lambda$ and $T$}\label{becproof}

In this section we are going to show the occurrence of BEC for small
$\lambda$ and low enough temperature. The main result is the following.

\begin{thm}[Existence of BEC]\label{thm1}
Let $E_\vecp=\sum_{i=1}^d (1-\cos(p_i))$ (where $p_i$ denotes the
components of $\vecp$) and
$$
c_d=\frac 1{(2\pi)^{d}} \int_{ [-\pi,\pi]^d} d\vecp\frac 1{E_\vecp} .
$$
In the thermodynamic limit,
\begin{eqnarray}\nonumber
&&\lim_{\Lambda\to \infty} \frac 1{|\Lambda|^2} \sum_{\vecx,\vecy\in\Lambda}
\gamma(\vecx,\vecy) \\
&&\geq \frac 12 -\frac12 \left(\half
\left[d(d+1)+4\lambda^2\right]^{1/2} c_d \right)^{1/2} - \frac
1{\beta} c_d, \label{beccurve}
\end{eqnarray}
with $\beta=1/(k_{\rm B}T)$ the inverse temperature. Moreover, if
$\varphi(\vecx)$ denotes the (normalized) eigenfunction corresponding
to the largest eigenvalue of $\gamma(\vecx,\vecy)$, then
$\lim_{\Lambda\to\infty} |\Lambda|^{-1} | \sum_\vecx \varphi(\vecx)|^2
= 1$, implying that the condensate wave function is constant in the
thermodynamic limit.
\end{thm}

Note that $c_d$ is finite for $d\geq 3$. Since the largest eigenvalue
of $\gamma(\vecx,\vecy)$ exceeds $|\Lambda|^{-1}
\sum_{\vecx,\vecy}\gamma(\vecx,\vecy)$, BEC
is proved if the right side of the expression (\ref{beccurve}) is
positive. This is in particular the case, for large enough $\beta$, as long as
$$
\lambda^2 < \frac 1{c_d^2}-\frac {d(d+1)}4.
$$
In $d=3$, $c_3\approx 0.505$ \cite{DLS}, and hence there is BEC for
$\lambda \lesssim 0.960$.  In \cite{DLS} it was also shown that $dc_d$
is monotone decreasing in $d$, which implies a similar result for all
$d>3$.

The main tool in our proof of Theorem~\ref{thm1} is an infrared bound
as in \cite{DLS}. The statement is as follows. For $A$ and $B$ bounded
linear operators, denote by
$$
(A,B)=\int _0^1 \Tr\left( A e^{-s \beta H} B e^{-(1-s)\beta H}
\right) ds / \Tr e^{-\beta H}
$$
the Duhamel two-point function.  For $\vecp\in\Lambda^*$ (the dual
lattice), $\vecp\neq \0$ and $\widetilde S^1_\vecp=|\Lambda|^{-1/2} \sum_\vecx
S_\vecx^1 \exp({\rm i} \vecp\cdot \vecx)$, we claim that
\begin{equation}\label{duhabound}
(\widetilde S_\vecp^1, \widetilde S^1_{-\vecp}) \leq \frac1{2 \beta E_\vecp}.
\end{equation}
The same is true with $\widetilde S^1_\vecp$ replaced by $\widetilde
S^2_\vecp$. This inequality  will allow us to prove the bound
(\ref{beccurve}) above.  Moreover, we will infer from
(\ref{duhabound}) the fact that there is only {\it one} large
eigenvalue (of order $|\Lambda|$) of the one-particle density matrix,
and the corresponding eigenfunction is constant (in the thermodynamic
limit).

We start by proving (\ref{duhabound}). The main ingredient is
{\it Gaussian domination}. More precisely, let $h$ be any real-valued
function on $\Lambda$, and
$$
Z(h)=\Tr \exp\left[ - \beta K(h) \right],
$$
where $K(h)$ is the modified Hamiltonian
\begin{eqnarray*}
      K(h)&=&\sum_{\langle \vecx\vecy\rangle} \left( \half
(S^1_\vecx-S^1_\vecy-h_\vecx+h_\vecy)^2 -
S^2_\vecx S^2_\vecy\right) \\ && +
      \lambda\sum_\vecx(-1)^\vecx S^3_\vecx.
\end{eqnarray*}
Note that for $h=0$ this operator agrees with $H$ up to a constant.

\begin{lem}\label{L1} For all real-valued functions $h$,
$$ Z(h)\leq Z(0).$$
\end{lem}

\begin{proof}
      We perform a unitary transformation that takes $S^2\mapsto -S^2$ and
      $S^3\mapsto -S^3$ on the B-sublattice. Since the trace does not
      change under unitary transformations, we have $Z(h)=\Tr \exp[ -
      \beta \widehat K(h) ]$ with
$$
\widehat K(h)= \sum_{\langle \vecx\vecy\rangle}\left( \half
(S^1_\vecx-S^1_\vecy-h_\vecx+h_\vecy)^2 +
      S^2_\vecx S^2_\vecy\right)+ \lambda\sum_\vecx S^3_\vecx.
    $$
    Compared to $K(h)$, the sign in front of the $S^2_\vecx S^2_\vecy$ term has
    changed, and the $(-1)^\vecx$ has vanished. The operator $\widehat K(h)$
    thus obtained is translation invariant. Since $S^1$ and $S^3$ are
    real self-adjoint matrices, and $S^2$ is imaginary and self-adjoint, we
    meet exactly the conditions for applying the result in
    \cite[Lemma~6.1 and proof of Theorem~4.2]{DLS} to prove the lemma.
\end{proof}

The infrared bound (\ref{duhabound}) follows from this lemma by using
the negativity of the second derivative,
$$
\left. \frac{d^2}{d\eps^2} Z(\eps h) \right|_{\eps=0} \leq 0.
$$
By performing the derivative, we obtain
$$
(A^\dagger,A)\leq \frac 1{\beta} \sum_{\langle \vecx\vecy\rangle}
|h_\vecx-h_\vecy|^2,
$$
where $A=\sum_{\langle \vecx\vecy\rangle}
(S^1_\vecx-S^1_\vecy)(h_\vecx-h_\vecy)$. We proved this inequality
only for real-valued $h$, in which case $A=A^\dagger$, but it
automatically extends in a standard way \cite{DLS} to complex-valued
$h$. In this case note that the adjoint of $A$ agrees with its complex
conjugate. Now, choosing $h_\vecx= \exp({\rm i} \vecp \cdot \vecx)$,
we obtain (\ref{duhabound}).  By invariance of the Hamiltonian under
rotations around the $S^3$-axis, the statement is also true with
$\widetilde S^1_\vecp$ replaced by $\widetilde S^2_\vecp$.

We now want to use \cite[Theorem~3.1]{DLS} to relate the Duhamel
two-point function to the ordinary thermal two-point function. For
that purpose, we have to evaluate the double commutators
\begin{eqnarray*}
     && [\widetilde S^1_\vecp,[H,\widetilde S^1_{-\vecp}]]+
[\widetilde S^2_\vecp,[H,\widetilde S^2_{-\vecp}]]\\ &&=-\frac 2 {|\Lambda|}
      \Big(H - \half \lambda|\Lambda| + 2\sum_{\langle
\vecx\vecy\rangle} S^3_\vecx
S^3_\vecy \cos \vecp\cdot(\vecx-\vecy)\Big).
\end{eqnarray*}
Let $C_\vecp$ denote the expectation value of this last expression,
$$
C_\vecp= \langle [\widetilde S^1_\vecp,[H,\widetilde
S^1_{-\vecp}]]+[\widetilde S^2_\vecp,[H,\widetilde S^2_{-\vecp}]]
\rangle\geq 0.
$$
The positivity of $C_\vecp$ can be seen from an
eigenfunction-expansion of the trace.  {F}rom \cite[Corollary~3.2 and
Theorem~3.2]{DLS} and (\ref{duhabound}) we infer that
\begin{equation}\label{dlsb}
\langle \widetilde S_\vecp^1 \widetilde S_{-\vecp}^1 + \widetilde S_\vecp^2
\widetilde S_{-\vecp}^2\rangle\leq \frac 12 \sqrt {
\frac  {C_\vecp}{E_\vecp}} \coth \sqrt{\beta^2 C_\vecp E_\vecp /4}.
\end{equation}
Using $\coth x \leq 1+1/x$ and Schwarz's inequality, we obtain for the
sum over all $\vecp\neq \0$,
\begin{eqnarray*}
     &&\sum_{\vecp\neq \0}\langle \widetilde S_\vecp^1 \widetilde S_{-\vecp}^1 +
     \widetilde S_\vecp^2 \widetilde S_{-\vecp}^2\rangle\\ &&\leq \frac 1{\beta}
     \sum_{\vecp\neq \0} \frac 1{E_\vecp} + \frac 12 \Big(
\sum_{\vecp\neq \0} \frac
     1{E_\vecp} \Big)^{1/2} \Big( \sum_{\vecp\neq \0} C_\vecp \Big)^{1/2}.
\end{eqnarray*}
We have $\sum_{\vecp\in\Lambda^*} C_\vecp = -2 \langle H
\rangle+\lambda|\Lambda|$, which can
be bounded from above using the following lemma. Its proof follows exactly the
same lines as \cite[Theorem~C.1]{DLS}.

\begin{lem}
The lowest eigenvalue of
\begin{equation}\label{sumt}
-\frac 12 S_\vecx^1 \sum_{i=1}^{2d} S_{\vecy_i}^1-\frac 12 S_\vecx^2
\sum_{i=1}^{2d} S_{\vecy_i}^2+ \lambda S_\vecx^3
\end{equation}
is given by $-\mbox{$\frac 14$}[d(d+1)+4\lambda^2]^{1/2}$.
\end{lem}

Since the Hamiltonian can be written as a sum of terms like
(\ref{sumt}), with $\vecy_i$ the nearest neighbors of $\vecx$, we get from
this lemma the lower bound
$$
H\geq -\frac {|\Lambda|}4\left
[d(d+1)+4\lambda^2\right]^{1/2}+\half\lambda|\Lambda|.
$$
With the aid of the sum rule
$$
\sum_{\vecp\in \Lambda^*}\langle \widetilde S_\vecp^1 \widetilde S_{-\vecp}^1 +
\widetilde S_\vecp^2
\widetilde S_{-\vecp}^2\rangle=\frac {|\Lambda|}2
$$
(which follows from $(S^1)^2=(S^2)^2=1/4$), we obtain a lower
bound, in the thermodynamic limit,
\begin{eqnarray}\nonumber
     &&\lim_{\Lambda\to \infty} \frac 1{|\Lambda|} \langle \widetilde S_\0^1
     \widetilde S_{\0}^1 + \widetilde S_\0^2 \widetilde S_{\0}^2\rangle\\
     &&\geq \frac 12 -\frac12 \left(\half
       \left[d(d+1)+4\lambda^2\right]^{1/2} c_d \right)^{1/2} - \frac
     1{\beta} c_d. \label{frrom}
\end{eqnarray}

The connection with Bose-Einstein condensation is as follows.
Since $H$ is real, also $\gamma(\vecx,\vecy)$ is real and we have
$$
\gamma(\vecx,\vecy)= \langle S_\vecx^+ S_\vecy^ - \rangle = \langle
S_\vecx^1 S_\vecy^1+S_\vecx^2 S_\vecy^2
\rangle.
$$
Hence, if
$\varphi_0=|\Lambda|^{-1/2}$ denotes the constant function,
$$
\langle \varphi_0 |\gamma| \varphi_0\rangle =  \langle \widetilde S_\0^1
     \widetilde S_{\0}^1 + \widetilde S_\0^2 \widetilde S_{\0}^2\rangle,
$$
and thus the bound (\ref{frrom}) implies (\ref{beccurve}).
In addition we claim that the infrared
bounds imply
$$
\langle \varphi |\gamma| \varphi \rangle \leq \const |\Lambda|^{2/d}
$$
for any normalized $\varphi$ that is orthogonal to $\varphi_0$,
with a constant that is independent of $\varphi$.  To see this,
consider the positive definite matrix $\langle \widetilde S^+_\vecp
\widetilde S^-_{-\vecq}\rangle$, with $\vecp\neq \0$, $\vecq \neq \0$.
The infrared bound (\ref{dlsb}) implies that the diagonal of this
matrix is bounded by $\langle \widetilde S^+_\vecp \widetilde
S^-_{-\vecp}\rangle\leq \const |\vecp|^{-2} \leq \const
|\Lambda|^{2/d}$ for $\vecp\neq \0$. Moreover, the matrix is almost
diagonal in the sense that $\langle \widetilde S^+_\vecp \widetilde
S^-_{-\vecq}\rangle\neq 0$ only if $q_i=p_i$ or $q_i=p_i \pm \pi$ (by
invariance under translation by two lattice sites). The largest
eigenvalue of such a matrix is bounded above by $2^d$ times the
maximum on the diagonal, namely $2^d \const |\Lambda|^{2/d} \ll
|\Lambda|$. This proves our claim.

We conclude that $\gamma(\vecx,\vecy)$ has exactly {\it one} large
eigenvalue, with corresponding eigenfunction equal to $\varphi_0$ as
$|\Lambda|\to \infty$. I.e., the condensate wave function is constant.
This is in contrast to the particle density, which shows the
staggering of the periodic potential. We show this in
Section~\ref{stagdens} below.  It also contrasts with the situation
for zero interparticle interaction, as discussed in
Section~\ref{sectfree}.

\section{No cusp, no gap}\label{sectgap}

The system is in a Mott insulator state at zero temperature if a
finite change in the chemical potential is required to change the
particle number in the ground state. We refer to this as a gap in the
chemical potential. More precisely, if $E_k$ denotes the lowest energy
of (\ref{ham}) restricted to the sector of $\half |\Lambda| +k$
particles (which corresponds to $S^3_{\rm tot}\equiv \sum_\vecx
S^3_\vecx = k$ in the spin language), a gap means that, for all $k$,
\begin{equation}\label{eq:gap}
E_{-k}+E_k-2E_0\geq c|k|
\end{equation}
for some $c>0$ independent of $\Lambda$ and $k$.
Note that particle-hole symmetry implies that
$E_{-k}+E_k-2E_0=2(E_{|k|}-E_0)$.

In the next section we prove
(\ref{eq:gap}) for sufficiently large $\lambda$. In this section we
will show that whenever there is BEC then (\ref{eq:gap}) fails.
In fact, we will prove that
\begin{equation}\label{nogap}
E_k-E_0 \leq \frac {c_k}{|\Lambda|}
\end{equation}
for some $k$-dependent $c_k>0$, which is independent of $\Lambda$,
i.e., that (\ref{eq:gap}) does not hold for finite $k$.

This, however, does not rule out the possibility that the macroscopic
system still acts as an insulator. To show that this is, indeed, also
not the case, we prove that (\ref{eq:gap}) fails for macroscopic $k$
as well.  More precisely, we will show that the thermodynamic limit of
the ground state energy per site, $e_\infty(\varrho)=\lim_{\Lambda\to \infty}
E_k /|\Lambda|$, where $k=(\varrho-\half)|\Lambda|$, satisfies a bound
\begin{equation}\label{nocusp}
0\leq e_\infty(\varrho)-e_\infty(\half)\leq \const (\varrho - \half)^2
\end{equation}
for $\varrho$ close to $1/2$, i.e., that there is no macroscopic cusp
in the energy at half-filling.

We will first prove (\ref{nocusp}). With $|0\rangle$ being the ground state
of $H$, and with $\vecy$ some point in the lattice, consider the states
$$|\psi_\vecy\rangle= e^{{\rm i}\eps S^2_{\rm tot}}
(S_\vecy^1+\half)|0\rangle.$$
The motivation is the following: we take
the ground state and first project onto a given direction of $S^1$ on
some site $\vecy$. If there is long-range order, this should imply
that essentially all the spins point in this direction now. Then we
rotate slightly around the $S^2$-axis. The particle number should then
go up by $\eps|\Lambda|$, but the energy only by $\eps^2|\Lambda|$.

The norm of $|\psi_\vecy\rangle$ is given by
$$\langle\psi_\vecy|\psi_\vecy\rangle=\langle
0|S_\vecy^1+\half|0\rangle=\half,$$
where we used the symmetry
$S^{1,2}\to -S^{1,2}$. We want to find an upper bound to the average
energy of these states, more precisely, $$\Delta E\equiv \frac
2{|\Lambda|} \sum_\vecy \langle\psi_\vecy|H-E_0|\psi_\vecy\rangle.$$
Here $E_0$ denotes the ground state energy of $H$, which is obtained
at half-filling (see Appendix~\ref{apphalf}). We claim that the
inequality
\begin{equation}\label{fircl}
e^{-{\rm i}\eps
     S^2_{\rm tot}} H e^{{\rm i}\eps S^2_{\rm tot}}\leq H + {\rm i}\eps [H,S^2_{\rm
     tot}] + \const \eps^2 |\Lambda|
\end{equation}
holds for some constant depending only on $d$ and $\lambda$. To see
this, consider, for self-adjoint operators $A$ and $C$,
$$
F_A(\eps)= e^{{\rm i}\eps C} A e^{-{\rm i}\eps C}.
$$
By Taylor's formula
\begin{equation}\label{taylor}
F_A(\eps) \leq F_A(0) + \eps F'_A(0) + \half \eps^2 \sup_{0\leq
\eta\leq \eps} \|F''_A(\eta)\|.
\end{equation}
Note that the last norm is actually independent of $\eta$, since
$e^{-{\rm i}\eps C}$ is unitary, and is given by the norm of the double
commutator $[C,[C,A]]$. After evaluating the double commutator for the case
in question, a simple bound gives (\ref{fircl}).

Consider now the first term on the right side of (\ref{fircl}). We obtain
\begin{eqnarray*}
&&\sum_\vecy \langle 0| (S_\vecy^1+\half) (H-E_0) (S_\vecy^1+\half) |0\rangle
\\ &&= \half \sum_\vecy \langle 0|
[S_\vecy^1,[H,S_\vecy^1]]|0\rangle\\ &&= \half
\Big\langle 0\Big| \sum_{\langle \vecx\vecy\rangle} 2 S_\vecx^2
S_\vecy^2 - \lambda \sum_\vecy
(-1)^\vecy S^3_\vecy \Big|0\Big\rangle \\&&= - \half \left( E_0 -
\half\lambda|\Lambda|\right),
\end{eqnarray*}
where we used rotational symmetry in the last step. The second term,
$$
\sum_\vecy \langle 0| (S_\vecy^1+\half) [H,S^2_{\rm tot} ]
(S_\vecy^1+\half)|0\rangle,
$$
is zero by symmetry, as can be seen in the following way. The
diagonal terms are zero by the symmetry $S_\vecx^{1,2}\to -
S_\vecx^{1,2}$ at all sites. The off-diagonal terms are $$\langle 0|
S^1_{\rm tot}[H,S^2_{\rm tot}] + [H,S^2_{\rm tot}] S^1_{\rm tot}
|0\rangle,$$
which is zero by the symmetry $S_\vecx^{1,3}\to
-S_\vecx^{1,3}$ at all sites, followed by a unit-vector translation in
any of the lattice directions. We therefore get that
$$\Delta E\leq  \const \eps^2
|\Lambda| + \frac {|E_0|+ \half\lambda|\Lambda|}{|\Lambda|}. $$

It remains to evaluate the average particle number of the states
considered. Using that $S^3_{\rm tot}|0\rangle=0$, we obtain
\begin{eqnarray*}
&&\frac 2{|\Lambda|} \sum_\vecy \langle 0| (S_\vecy^1+\half) e^{-{\rm i}\eps
S^2_{\rm tot}} S^3_{\rm tot} e^{{\rm i}\eps S^2_{\rm
tot}}(S_\vecy^1+\half)|0\rangle \\&&=\frac 2{|\Lambda|} \sin\eps \big\langle
0\big| \left( S^1_{\rm tot}\right)^2\big|0\big\rangle,
\end{eqnarray*}
which is of the order $\eps|\Lambda|$ if there is BEC.  Choosing
$\eps$ proportional to the chemical potential $\mu$, we obtain as an
upper bound for the ground state energy of $H-\mu S^3_{\rm tot}$ $$
E_0\left(1-|\Lambda|^{-1}\right)+\half\lambda - c \mu^2 |\Lambda| $$
for small $\mu$, with $c>0$ in the case of BEC. By taking a Legendre
transform, we arrive at (\ref{nocusp}).

To prove (\ref{nogap}) we use as a trial state $(S_{\rm tot}^+)^k
|0\rangle$, with $k\geq 1$.  Using the particle-hole symmetry of $H$ as well as
the fact that we are considering the ground state, we get the bound
$$
E_k \leq E_0+ \frac 12 \frac {\langle [ (S_{\rm tot}^-)^k, [ H, (S_{\rm
tot} ^+)^k]] \rangle }{\langle (S_{\rm tot}^-)^k (S_{\rm tot}^+)^k
\rangle }.
$$
Since there is BEC, $\langle (S_{\rm tot}^-)^k (S_{\rm tot}^+)^k
\rangle\geq c_k' |\Lambda|^{2k}$ for some $c_k'>0$ modulo lower order
terms as $|\Lambda|\to \infty$. All we have to show is that $[(S_{\rm
    tot}^-)^k, [ H, (S_{\rm tot}^+)^k]] \leq c_k'' |\Lambda|^{2k-1}$ for
some constant $c_k''$. This is clear, however, since altogether there
are $|\Lambda|^{2k+1}$ factors, and the 2 commutators reduce the power
by two. Hence we obtain (\ref{nogap}).

\section{Gap for large $\lambda$, and absence of BEC for
large $\lambda$ or high temperature}\label{loops}

We shall now present explicit bounds for a region of values of
$(\lambda,T)$ for which BEC is absent. This region includes:
\begin{enumerate}
\item [(i)]  all  $\lambda \ge 0$ at
$ k_{\rm B} T> d/(2 \ln 2)$,
\item [(ii)] all $T\ge 0$ at
$\lambda \ge 0$  such that $\lambda + |e(\lambda)|> d$.
\end{enumerate}
The absolute (i.e., without specifying $N$) ground state energy per
site for a finite $\Lambda$, which is always obtained at half-filling,
is denoted by $e(\lambda)$. Note that $e(\lambda)<0$.

In this regime, the particles are localized in the sense that the
transition amplitudes decay exponentially.  Short excursions occur
locally in ``space-time'', however a long distance transition requires
a linked chain, or percolation, of such local events and the amplitude
for that decays exponentially, as in sub-critical percolation models.
One may discern here two distinct mechanisms contributing to the
localization: at high $\lambda$ localization is caused by the
confinement due to the staggered structure of the potential, whereas
at high temperatures it is a combined effect of the exclusion (no more
than one particle at a lattice site) with the reduced amplitude for
coordinated moves by neighboring particle.
The above picture is made precise in a representation of the matrix
elements of $e^{-\beta H}$ which in effect involves an ``imaginary time''.

A more inclusive statement of the condition under which our results
hold is
\begin{equation}    \label{eq:highLT}
\beta > \frac{-1}{\lambda - f} \ln \left (1-\frac{\lambda -f}{d}\right)  ,
\end{equation}
where $f \equiv f(\beta, \lambda) = -(\beta |\Lambda|)^{-1} \ln \Tr
e^{-\beta H}$ is the free energy per site, which satisfies
$$
-f(\beta, \lambda)  \geq  \max\{|e(\lambda)|, \beta^{-1} \ln
2-\half\lambda\} \, .
$$
Under condition (\ref{eq:highLT}), we define $\nu>0$ by
\begin{equation} \nonumber
    e^{-\nu} \equiv \ \frac
d{\lambda-f}\left(1-e^{-\beta(\lambda -f)}\right) < 1   .
\end{equation}

\begin{thm}[Mott insulator phase] \label{thm:noBEC}
   Throughout the regime where (\ref{eq:highLT}) holds the thermal
   average two-point function decays exponentially. More specifically,
   for any $\xi <\nu$
\begin{equation}\label{expdec}
\gamma(\vecx,\vecy)= \frac{\Tr  a^\dagger_\vecx
a^{\phantom\dagger}_\vecy e^{-\beta H}}{\Tr
e^{-\beta H}} \, \le \,   C_\xi \, e^{-\xi |\vecx-\vecy|}
\end{equation}
with $C_\xi = [1-e^{\xi -\nu }]^{-1}$.
Similar decay also holds in the finite volume ground state
(corresponding to the limit
$\beta \to \infty$ in (\ref{expdec})).

Moreover, the ground state energy $E_{k}$ for
particle number $\half|\Lambda|+k$ satisfies
\begin{equation}\label{finitegap} E_k+E_{-k}-2E_0=2(E_k-E_0)\, \geq\,
   c \,|k|
\end{equation}
with $c=2(\lambda+|e(\lambda)|-d)$, which is strictly positive for
large $\lambda$ (independently of the volume $|\Lambda|$).
\end{thm}

Eq.\ (\ref {finitegap}) is, in fact, the consequence of a more
explicit result: If $\p_k$ denotes the projection onto the subspace of
fixed particle number $N= \half|\Lambda|+k$, then
\begin{equation}  \label{eq:gapcondition}
\frac{\Tr \p_k e^{-\beta H} }{\Tr \p_0 e^{-\beta H}} \leq 2\,
e^{-\alpha\beta|k|}
\left[ \frac{e^{2}}{1-B(\alpha)}
\frac{|\Lambda |}{|k|} \right]^{|k|}  ,
\end{equation}
for any $\alpha>0$ for which
$$
B(\alpha)  \equiv  d \int^\beta_0 e^{-(\lambda
   +|f_0(\beta,\lambda)|-\alpha)t}  dt  <  1\, ,
$$
with $f_0(\beta,\lambda)$ denoting the free energy for the system
with fixed particle number $N=|\Lambda|/2$.  Eq.\ (\ref{finitegap}) is
derived by considering the leading terms in (\ref{eq:gapcondition}) in
the limit $\beta \to \infty$ at fixed $\Lambda$.  In the thermodynamic
limit Eq.\ (\ref {finitegap}) means that, in contrast to Eqs.\
(\ref{nogap}) and (\ref {nocusp}), the energy per site,
$e_\infty(\varrho)$, has a cusp at $\varrho=1/2$, and hence (by
Legendre transform) the half filled state $\varrho=1/2$ corresponds to
a whole interval of values for the chemical potential.

For non-zero temperature the energy dependence on $k$ may show some
rounding due to thermal excitations, however there is a cusp in the
energy per site, when this function is viewed on a scale in which
$|k|/|\Lambda| \geq e^{-r \beta c}$, with some $r<1$.

%%%%%%%

Theorem~\ref{thm:noBEC} is derived using a representation for the
matrix elements of the relevant operators in the basis that
diagonalizes $\{S^3_\vecx\}$.  An important fact here is that the
matrix elements of $e^{-\beta H}$ are positive in this basis.  The
natural expansion for these matrix elements, e.g., via the Lie-Trotter
formula, yields a functional-integral representation in terms of
integrals over ``space-time'' configurations, which are represented
below by $\omega$.
The resulting functional integral is not only
positive, but also reflection positive, and we make use of that fact.
However, reflection positivity is not essential for the qualitative
picture, as even without it we would obtain similar results with only 
slightly weaker bounds -- with $f$ replaced by $0$.

The measure space over which the integration takes place is
a Cartesian product of the set of initial spin configurations times the
space of configurations of ``rungs'', linking pairs of neighboring
``space time'' sites.  A rung is parametrized  by a pair
$\{\vecx,\vecy\}$ of neighboring lattice sites and $t\in [0,\beta]$.
For the matrix elements of $e^{-\beta H}$ between states which
correspond to a pair of specified spin configurations, one naturally
finds an integral over configurations of arbitrary number of
rungs, over which we integrate with
an ``ideal gas''-like measure, in which
$n$-tuples  are summed and integrated over with the weights
$\frac{z^n}{n!}\, dt_1 \cdots dt_n$.  Each rung represents a 
transformation of the spin configuration affected by a specific term 
in the Hamiltonian, and the fugacity-like
parameter $z$ is the corresponding amplitude, which in the case of 
the Hamiltonian considered here is $z=1/2$.

%%%%%%%
It is particularly convenient to express the spin, or particle,
configuration in terms of the time-lines of the `quasi-particles'
which are defined through the occupation numbers
$n_\vecx=\frac{1}{2}+(-1)^\vecx S^3_\vecx$.  There are no
quasi-particles in the configuration that minimizes the potential
energy, i.e, if there are $\half |\Lambda|$ particles that sit on the
B-sublattice.  The presence of a quasi-particle means either the
presence of a particle ($S^3=+1/2$), if the site is even
(A-sublattice), or the absence of one ($S^3=-1/2$), if the site is odd
(B-sublattice).  It is easy to check that in this representation the
operators $a_\vecx^\dagger$ and $a^{\phantom\dagger}_\vecx$ act as
insertion of a source and, correspondingly, sink of excess spin
(relative to the potential-minimizing configuration), although the
direction in which the excess spin propagates changes with the parity
of $\vecx$. Namely, creation of a particle on an A site (or
annihilation on a B site) results in a quasiparticle running \lq
upward in time\rq, whereas a quasiparticle running \lq downward in
time\rq\ originates from annihilation of a particle on an A site (or
creation on a B site).

\begin{figure}[htf]
\includegraphics[width=8.04cm, height=4.22cm]{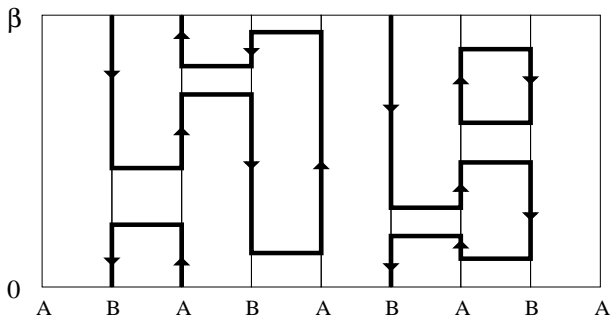}
\caption{Loop gas describing paths of quasi-particles for particle
    number $N=|\Lambda|/2-1$. A line on an A site means presence of a
    particle, while on a B site it means absence. The horizontal rungs
    correspond to hopping of a particle.}
\label{loopfig}
\end{figure}

Proceeding along the above lines, as  explained in greater detail in
\cite{AN}, one obtains
\begin{equation} \label{eq:nosources}
\Tr  \p_k e^{-\beta H}  = \int \upsilon_{1/2}(d\omega)
e^{-\lambda |\omega|}  I[\nu(\omega)=k] \, ,
\end{equation}
where $\omega$ represents a configuration of a family of disjoint
oriented loops in $\Lambda \times [0,\beta]$, defined with periodic
boundary conditions in `time' ($[0,\beta]$), whose orientation
alternates with $\vecx$, being `up' along A sites and `down' along B
sites.  For each configuration, $|\omega|$ is the total `vertical'
length of the time lines in $\omega$, and $\nu(\omega)$ is the total
winding number in the periodic `time' direction. The indicator
function $I[\nu(\omega)=k]$ is 1 if the loop configuration $\omega$
has total winding number $k$, and 0 otherwise. The winding number can
also be computed by adding the spin orientations of the sites occupied
by quasi-particles, along any `constant time' cut through the diagram.
The measure $\upsilon_z(d \omega)$ corresponds to integration, with
weights $z dt$, over the times at which the jumps to neighboring
lattice sites occur, and summation over the possible numbers of such
jumps.  In effect, as mentioned above, the integral is over an `ideal gas'-like
distribution of the horizontal rungs in the diagram depicted in
Figure~\ref{loopfig} with the fugacity parameter taking here the 
value $z=1/2$.  That value is
dictated by the Hamiltonian, where one finds $1/2$ in front of the 
\lq hopping term\rq\ $S^+_\vecx S^-_\vecy$.  For later use, we find 
it convenient
to consider the measures $\upsilon_z(d\omega)$ for general $z>0$, not
only $z=1/2$.

Likewise, for $\vecx\neq \vecy$,
\begin{eqnarray} \label{eq:sources}
&&\Tr  a^\dagger_\vecx a^{\phantom\dagger}_\vecy \p_k e^{-\beta H} \\
&& =
\int_{\partial \omega = \delta_{(\vecx,0)} - \delta_{(\vecy,0)} }\
       \upsilon_{1/2}(d\omega) \,  e^{-\lambda |\omega|}  \, I[\nu(\omega)=k]
\nonumber
\end{eqnarray}
where $\partial \omega$ is the set of sources of $\omega$. More
precisely, the configurations that contribute to the last integral
have exactly {\em one} curve starting at $\vecx$ and ending at $\vecy$, both at
time 0, for which we shall use the symbol $\gamma$, and otherwise only
closed loops, of the kind which appear in the corresponding trace
without the source/sink operators.

Denoting by $\Aa^{\rm o}$ the set of configurations $\omega$ that
consist only of loops, i.e., closed curves, and by
$\Aa^{(\vecx,\vecy)}$ the set of
configurations containing one curve connecting $\vecx$ with $\vecy$ (at time
$0$) and otherwise only closed curves, the above representation
yields for the thermal average unconstrained by the particle number
\begin{equation}\label{eq:pathrep}
\langle S_\vecx^+ S_\vecy^- \rangle = \frac {\int_{\Aa^{(\vecx,\vecy)}}
\upsilon_{1/2} (d\omega) e^{-\lambda |\omega|}}{\int_{\Aa^{\rm o}}
\upsilon_{1/2} (d\omega) e^{-\lambda |\omega|}} \, .
\end{equation}

Now let $\Bb^{(\vecx,\vecy)}\subset\Aa^{(\vecx,\vecy)}$ denote the set
of $\omega$'s which consist of exactly {\it one} curve connecting
$\vecx$ and $\vecy$, no other curves.  For a collection of disjoint curves $\{ \gamma_j
\}$, let $\Aa_{\{ \gamma_j \}} ^{\rm o}\subset \Aa^{\rm o}$ denote the set of
$\omega$'s that avoid the collection $\{ \gamma_j\}$ or, in other
words, that are consistent with it in the sense that insertion of the
curves into $\omega$ would still give an admissible configuration of
non-intersecting curves. The measures $\upsilon_z(d\omega)$ obviously
have the product structure
\begin{eqnarray*}
&&\int _{\Aa^{(\vecx,\vecy)}} \upsilon_{z} (d\omega) e^{-\lambda
|\omega|} \\&& = \int
_{\Bb^{(\vecx,\vecy)}} \upsilon_{z} (d\gamma) e^{-\lambda |\gamma|} \int
_{\Aa^{\rm o}_\gamma} \upsilon_{z} (d\omega) e^{-\lambda |\omega|}   ,
\end{eqnarray*}
where we identified $\omega$ with $\gamma$ in the first integral on
the right.  The following is a convenient bound on the last factor.

\begin{lem}[Contour bound] \label{lem:rp}
For any family of disjoint curves $\{\gamma_j \}$,
\begin{equation}\label{claim2}
\int _{\Aa^{\rm o}_{\{\gamma_j \}}} \upsilon_{z} (d \omega) \, e^{-\lambda
|\omega|} \,
      \leq \, e^{\sum_j |\gamma_j| f}  \int _{\Aa^{\rm o}}
\upsilon_{z} (d  \omega) \,
e^{-\lambda |\omega|} ,
\end{equation}
where $f$ is the free energy per site.
Furthermore, a similar bound holds for the integrals further restricted
by the condition $I[\nu(\omega)=0]$ and $f$
replaced by $f_0$ (which equals $ -(\beta |\Lambda|)^{-1} \ln \Tr \p_0\,
e^{-\beta H}$).
\end{lem}

The proof, which is presented in Appendix~\ref{app:pathRP}, uses
the chessboard inequality, which is a consequence of the reflection positivity
of the functional integral.
As was mentioned already, our main qualitative conclusions do not 
require this result, and would follow already from the trivial bound
in which $f$ and $f_0$, which satisfy $f < f_0 < 0$, are replaced by $0$.

Lemma~\ref{lem:rp} implies that
\begin{equation}\label{gamma}
\langle S^+_\vecx S^-_\vecy\rangle \leq \int _{\Bb^{(\vecx,\vecy)}}
\upsilon_{1/2}(d\gamma)
e^{-|\gamma|\left(\lambda- f\right)}.
\end{equation}
To evaluate such expressions, it is useful to consider the quantity
$$
\chi (z,\lambda) = \sup_{\hat x,\hat y} \int_{\Bb^{(\hat x,\hat
      y)}} \upsilon_z(d \gamma) e^{- \lambda |\gamma|},
$$
where $\hat x=(\vecx,t_x)$ and $\hat y=(\vecy,t_y)$ are arbitrary
points in $\Lambda \times [0,\beta]$.

\begin{lem}\label{lem:chibounds}
If $2 z d \int_0^{\beta} e^{-\lambda t} dt <1$, then
\begin{equation}  \label{eq:chibound}
\chi (z,\lambda) \leq  \frac {1}{1-2 z d \int_0^{\beta}
e^{-\lambda t} dt} .
\end{equation}
Furthermore, for any $\{\hat x,\hat y\}$ and $\xi, \alpha>0$,
\begin{eqnarray}
&&\int_{\Bb^{(\hat x,\hat y)}} \upsilon_z(d \gamma)  e^{- \lambda
|\gamma|}  \le  e^{-\xi|\vecx-\vecy|}   \chi (z e^{\xi},\lambda) ,
\nonumber \\ \mbox{  }
\label{eq:decay} \\
   &&\int_{\Bb^{(\hat x,\hat y)}} \upsilon_z(d \gamma) e^{- \lambda
|\gamma|}  I[|\gamma|\ge t] \le e^{-\alpha t}   \chi (z,\lambda-\alpha ) .
\nonumber
\end{eqnarray}
\end{lem}

\begin{proof}
   The first inequality here is a random walk bound, which is derived
   by the following \lq renewal-type\rq argument: split the integral
   into a part that comes from curves that do not jump at all (which
   only occurs if $\vecx=\vecy$) and a part where $\gamma$ has at least
   one jump. The contribution from the path that does not jump is at
   most 1.  The first jump can be in $2d$ possible directions, hence
   one gets
$$\chi \le  1 + 2 z d \int_0^\beta e^{-\lambda t} dt  \,  \chi \, .
$$
Since in a finite volume
$\chi<\infty$ {\em a priori}, this yields (\ref{eq:chibound}).

The rest follows by fairly direct arguments, noting that
$\upsilon_{ze^\xi} (d\gamma)= \upsilon_z(d\gamma) e^{\xi \#(\gamma)}$,
where $\#(\gamma)$ denotes the number of jumps of $\gamma$, which is
greater or equal to $|\vecx-\vecy|$ in the case in question. The second
inequality in (\ref{eq:decay}) is obtained by estimating
$I[|\gamma|\geq t]\leq e^{\alpha(|\gamma|-t)}$ for any positive
$\alpha$.
\end{proof}

We shall now use the above functional representation
to derive Theorem~\ref{thm:noBEC}.

By applying the bounds (\ref{gamma}), (\ref{eq:chibound}) and
(\ref{eq:decay}) to the random walk representation (\ref{eq:pathrep}),
we see that under the condition stated in (\ref{eq:highLT}) the
two-point function decays exponentially, as claimed in (\ref{expdec}).

To prove (\ref{eq:gapcondition}) we start by noting that in case
$N\neq |\Lambda|/2$ the integral in (\ref{eq:nosources}) is over
configurations with a non-trivial winding number, $\nu(\omega)=k$.
Each such configuration includes a collection of `non-contractible'
loops $\{\gamma_j\}$ with non-zero winding numbers,
$\nu_j\equiv\nu(\gamma_j)\neq 0$. The total length of the set
$\{\gamma_j\}$ is at least $\beta |k|$.  We shall bound
the relative weight of such configurations by using the second bound in
(\ref{eq:decay}) and combining it with an argument whose purpose is to
control the `entropy' of such a collection of long loops.

Each non-contractible loop $\gamma_j$ can be labeled by a starting
point $\vecx_j\in \Lambda$ where $\gamma_j$ crosses the 0 time line,
and a winding number $\nu_j$.  We shall actually overcount by summing
over all possible $\vecx_j$'s as starting points for the loops (with
the only restriction that $\vecx_j\neq \vecx_i$ for $j\neq i$), and
over all possible winding numbers $\nu_j$ with $\sum_j |\nu_j| \geq
|k|$.

For a given collection $\{\gamma_j\}$ of
non-contractible loops, we can bound the integral over the
remaining loops by an integral over loops with zero total winding
number that avoid the $\gamma_i$'s. Hence, starting from
(\ref{eq:nosources}), we get the bound
\begin{eqnarray*}
&&\frac{\Tr\p_k e^{-\beta H}}{\Tr\p_0 e^{-\beta H}}
\leq \sum_{\{\vecx_1,\vecx_2,\dots\}\subset \Lambda} \\  &&
\prod_{j}  \int_{\Bb^{(\vecx_j,\vecx_j)}} \upsilon_{1/2}(d\gamma_j)
e^{-\lambda|\gamma_j|} I[ \nu(\gamma_j)\neq 0] \\ &&
\times  I \big[ \mbox{$\sum_j$} |\nu(\gamma_j)| \geq |k|\big]
\, I_{\rm ni}[\{\gamma_1,\gamma_2,\dots\}] \\ &&
\times  \frac{ \int
_{\Aa^{\rm o}_{\{\gamma_j\} }} \upsilon_{1/2} (d\omega)
e^{-\lambda |\omega|}  I\left[ \nu(\omega) =0 \right] }
{ \int_{\Aa^{\rm o}} \upsilon_{1/2} (d\omega)  e^{-\lambda |\omega|}  I\left[
\nu(\omega) =0 \right]}.
\end{eqnarray*}
Here $I_{\rm ni}$ denotes the indicator function for having only {\it
    non-intersecting} loops.  Using the chessboard bound of
Lemma~\ref{lem:rp}, the last fraction can be bounded above by $\prod_j
e^{f_0(\beta,\lambda) |\gamma_j|}$, with $f_0(\beta,\lambda)<0$ the free
energy per site at $N=|\Lambda|/2$.  Applying the bounds of
Lemma~\ref{lem:chibounds} to
the integral over $n$ loops $\gamma_i$ with given absolute value of
the winding number, $m_i=|\nu_i|$ , we have, for any $\alpha >0$,
\begin{eqnarray}\nonumber
&& \prod_{j=1}^n  \int_{\Bb^{(\vecx_j ,\vecx_j
)} } \upsilon_{1/2}(d\gamma_j) e^{-(\lambda+|f_0|)
|\gamma_j|}I[|\nu(\gamma_j)|=m_j ]
\\ && \leq  \widehat \chi (\alpha)^n e^{-\alpha \beta \sum_j m_j }. 
\label{summ}
\end{eqnarray}
Here $\widehat \chi (\alpha) =
\chi(1/2,\lambda+|f_0(\beta,\lambda)|-\alpha)$, which is finite if
$\alpha$ is not too large.

To complete the bound, we have to sum the right side of (\ref{summ})
over all the
possible choices of the collection of the starting points of the
winding loops
$\{ \vecx_1, \dots, \vecx_n\} $, and over all
possible winding numbers $\nu_j$ with $|\nu_j|>1$ and
$\sum_j |\nu_j|\geq |k|$.  To do so, we employ the following
device.
Defining
$$
P(z)= 1+  \widehat\chi(\alpha)  \sum_{i=1}^{|\Lambda|/2} (z \delta )^i ,
$$
with $\delta=e^{-\alpha\beta}$, we see that the sum in question is
given by the sum of all the coefficients of $z^l$ in
$P(z)^{|\Lambda|}$ with powers $l\geq |k|$.  Hence
$$
\frac{\Tr \p_k e^{-\beta H}}
{\Tr \p_0 e^{-\beta H}}  \leq  \frac{ 1}{2\pi {\rm i}} \oint_{|z|=R}
\frac{d z}{z^{|k|+1}} \frac{1}{1-z^{-1}}  P(z)^{|\Lambda|},
$$
where $R$ can be any number greater than $1$.
The contour integral serves as a filter, selecting for us the
relevant coefficients of the polynomial $P(z)^{|\Lambda|}$.
A simple bound shows that
the above quantity is bounded from above by
$$
\inf_{1< R < 1/\delta} \frac{1 }{1-R^{-1} } \,
\frac{e^{|\Lambda|\widehat \chi(\alpha) \frac { R\delta}{1-R\delta}}
}{R^{|k|} }.
$$
We now choose $R= k/(\delta\widehat\chi(\alpha)|\Lambda|)$,
assuming that this quantity is greater or equal to~2. Note that
$\delta R \leq \half$, since $|k|\leq \half|\Lambda|$ and
$\widehat\chi(\alpha)\geq 1$. Hence we obtain
$$
\frac{\Tr \p_k e^{-\beta H}}
{\Tr \p_0 e^{-\beta H}} \le
2 \, e^{-\alpha \beta  |k| }
\left[ e^2 \widehat \chi(\alpha)  \frac{|\Lambda |}{|k|} \right]^{|k|} .
$$
This inequality is also valid, however, if $|k|/(\delta\widehat\chi(\alpha)
|\Lambda|)<2$, since the resulting bound then exceeds $1$, which is greater
than the left hand side (as shown in Appendix~\ref{apphalf}).  This proves the
claim made in (\ref{eq:gapcondition}), which presents sufficient conditions
for the existence of a cusp in the energy dependence on $N$, i.e., of
a gap in the chemical potential.

\section{Non-constancy of the density}\label{stagdens}

In Section~\ref{becproof} above we have demonstrated the existence of
BEC for small $\lambda$ and $T$, and also that the condensate wave
function is constant. Despite this fact the particle density has the
periodicity of the external potential and is not constant for
$\lambda\neq 0$. More precisely, the following result is proved below.

\begin{thm}[Non-constancy of the density]
   Let $\varrho(\vecx)=\gamma(\vecx,\vecx)$ denote the particle density
   in the thermal equilibrium state at inverse temperature $\beta$.
   With $e(\lambda,\beta)= |\Lambda|^{-1} \langle H \rangle$ equal to
   the energy per site,
\begin{equation}\label{densstag}
\frac 1{|\Lambda|} \left| \sum_{\vecx\in \Lambda} (-1)^\vecx
\varrho(\vecx)\right|
\geq \frac { \lambda\,| e(0,\beta)|^2} { 2 d^2(3d+\lambda)}.
\end{equation}
\end{thm}

We first prove this for the
ground state. We will show that the ground state energy of $H$,
denoted by $E(\lambda)$, satisfies
\begin{equation}\label{enes}
E(\lambda)\leq E(0) + \half\lambda |\Lambda| - c \lambda^2 |\Lambda|,
\end{equation}
with $c=\half e_0^2 d^{-2}/(3d+\lambda)$. (Here $e_0=E(0)/|\Lambda|$
denotes the ground state energy per site at $\lambda=0$.) Eq.
(\ref{enes}) implies (\ref{densstag}) by the following argument.
Write $H= H_0 + \lambda W$, with the obvious notation for $H_0$ and
$W$.  Since $E(\lambda)$ is a concave function of $\lambda$, we have
$$
\langle W\rangle = E'(\lambda) \leq \frac {E(\lambda)-E(0)}\lambda
\leq \half |\Lambda| -  c \lambda |\Lambda|.
$$
On the other hand,
$$
\langle W\rangle = \sum_{\vecx\in \Lambda}\left[\half+ (-1)^\vecx \langle
S_\vecx^3\rangle\right]
= \half|\Lambda|+ \sum_{\vecx\in \Lambda} (-1)^\vecx \varrho(\vecx).
$$
Combining the last two equations, we obtain
$$
\frac 1{|\Lambda|} \left| \sum_{\vecx\in \Lambda} (-1)^\vecx
\varrho(\vecx)\right|
\geq c\lambda,
$$
which proves our claim.

It remains to show (\ref{enes}). To do this, let the operator $C$ be given by
$$
C=\half \sum_{\langle \vecx\vecy\rangle} (-1)^\vecx \left(
    S_\vecx^1 S_\vecy^2 - S_\vecx^2 S_\vecy^1\right).
$$
As in the Hamiltonian, the sum is over all nearest neighbor pairs,
each pair counted only once. This operator has the nice property that
$$
[C,W]={\rm i} H_0.
$$
Proceeding as in (\ref{fircl})--(\ref{taylor}), a simple bound of
the relevant double commutators gives
$$
e^{{\rm i}\eps C} W e^{-{\rm i}\eps C} \leq  W - \eps H_0 + \eps^2
\frac{d^2}2 |\Lambda|
$$
as well as
$$
e^{{\rm i}\eps C} H_0 e^{-{\rm i}\eps C} \leq H_0 + {\rm i}\eps [C,H_0] +
\eps^2\frac{3d^3}{2} |\Lambda|.
$$
Note that the $\eps^2$-terms are of order of the volume, due to the
fact that both $C$ and $H$ contain only nearest neighbor terms.  With
$|0\rangle$ the ground state of $H_0$, we therefore have, using
$\langle 0 | W|0\rangle=\half |\Lambda|$ and
$\langle 0|[C,H_0]|0\rangle=0$,
\begin{eqnarray*}
E(\lambda)&\leq& \langle 0| e^{{\rm i}\eps C} H e^{-{\rm i}\eps C} |
0\rangle \\ &\leq& E(0)(1-\eps \lambda) +\half\lambda|\Lambda|+
\half \eps^2 d^2
(3d + \lambda) |\Lambda|.
\end{eqnarray*}
Now the optimal choice of $\eps$ is $\eps=2 c \lambda/e_0$, which
finishes the proof of (\ref{enes}).

A similar argument works at positive temperature. It shows that the
free energy depends non-trivially on $\lambda$, and by the same
concavity argument as above this implies the non-constancy of the
density, also at positive temperature, as claimed in (\ref{densstag}).

\section{The non-interacting gas}\label{sectfree}

The interparticle interaction is essential for the existence of a Mott
insulator phase for large $\lambda$. In case of absence of the
hard-core interaction, there is BEC for any density and any $\lambda$
at low enough temperature (for $d\geq 3$). To see this, we have to
calculate the spectrum of the one-particle Hamiltonian $-\half\Delta +
V(\vecx)$, where $\Delta$ denotes the discrete Laplacian and
$V(\vecx)=\lambda (-1)^\vecx$. The spectrum can be easily obtained by
noting that $V$ anticommutes with the off-diagonal part of the
Laplacian, i.e., $\{ V, \Delta+2d\} = 0$. Hence
$$
\left(-\half\Delta - d + V(\vecx) \right)^2 = \left(-\half \Delta -
d\right)^2 + \lambda^2,
$$
so the spectrum is given by
$$
d\pm \sqrt{\left(\mbox{$\sum_i$} \cos p_i\right)^2 +\lambda^2},
$$
where $\vecp\in\Lambda^*$. In particular, $E(\vecp)-E(0)\sim \half d
(d^2+\lambda^2)^{-1/2} |\vecp|^2$ for small $|\vecp|$, and hence
there is BEC for
low enough temperature. Note that the condensate wave function is of
course {\it not} constant in this case, but rather given by the eigenfunction corresponding to the lowest eigenvalue of $-\half\Delta+\lambda(-1)^\vecx$.  

\section{Conclusion}

We have introduced a lattice model, which is similar to the usual
Bose-Hubbard model and which describes the transition between
Bose-Einstein condensation and a Mott insulator state as the strength
$\lambda$ of the optical lattice potential is increased.  While the
model is not soluble in the usual sense, we can prove rigorously all
the essential features that are observed experimentally.  These
include the existence of BEC for small $\lambda$ and its suppression
for large $\lambda$, which is a localization phenomenon depending
heavily on the fact that the Bose particles interact with each other.
In the Mott insulator regime we prove the existence of a gap in the
chemical potential, which does not exist in the BEC phase and for
which the interaction is also essential. Bounds on the critical
$\lambda$ as a function of temperature are included.

\begin{acknowledgments}
    We are grateful to Letizzia Wastavino for help with
    Figures~\ref{fig1} and~\ref{loopfig}.
The work was supported in part by the NSF grants
PHY 9971149 (MA), PHY 0139984-A01 (EHL), and DMS-0111298 (JPS); by
EU grant HPRN-CT-2002-00277 (JPS and JY),
by MaPhySto -- A Network in Mathematical Physics and
Stochastics funded by The Danish National Research Foundation (JPS),
and by grants from the Danish research council (JPS).
\end{acknowledgments}

\appendix

\section{Half-filling and reflection positivity}\label{apphalf}

In this appendix we will show that $H$ has a {\it unique} ground state
which has particle number $|\Lambda|/2$. For $\lambda=0$, this was
previously shown in \cite{Mattis}.  We also establish the
corresponding result at positive temperature, namely that the
canonical partition function is maximal for particle number
$N=|\Lambda|/2$, although we do not prove that the maximum is obtained
only at half-filling.

The operator $H$ commutes with $\sum_\vecx S^3_\vecx$. By a Perron-Frobenius
argument the ground state of $H$ restricted to the subspace with fixed
value of $\sum_\vecx S^3_\vecx$ is unique.  We claim that the absolute ground
state of $H$ corresponds to the value $\sum_\vecx S^3_\vecx=0$.  To prove this
we shall use reflection positivity.

We divide the lattice in a left part and a right part
$\Lambda=\Lambda_L\cup\Lambda_R$ of equal size.  We shall identify the
space ${\cal H}_1=\bigotimes_{\vecx\in\Lambda_L}\C^2$ with the
space $\bigotimes_{\vecx\in\Lambda_R}\C^2$, by identifying factors
reflected in the middle plane. We may therefore write the total Hilbert
space as ${\cal H}={\cal H}_1\otimes{\cal H}_1$.  We may then write
$$
H=H_L\otimes I+I\otimes H_R-\half \sum_{\langle \vecx\vecy\rangle\in
M}(S^+_\vecx S^-_\vecy +S^-_\vecx S^+_\vecy),
$$
where $H_L$ and $H_R$ act on ${\cal H}_1$ and $M$ denotes the set
of bonds going from the left sublattice to the right sublattice (note
that because of the periodic boundary condition these include the
bonds that connect the right boundary with the left boundary). Note
that $H_L\ne H_R$.

We now change $H$ to the unitarily equivalent operator $H'$ for which
at all sites on the right sublattice we change $S_\vecx^\pm\to S_\vecx^\mp$
and $S_\vecx^3\to-S_\vecx^3$.  We have
$$
H'=H_L\otimes I+I\otimes H_L-\half \sum_{\langle \vecx\vecy\rangle\in
M}(S^+_\vecx S^+_\vecy+S^-_\vecx S^-_\vecy).
$$
The same unitary will change $\sum_\vecx S^3_\vecx$ to
$S'=S\otimes I-I\otimes S$,
where $S=\sum_{\vecx\in\Lambda_L}S^3_\vecx$ acts on ${\cal H}_1$.

Let $|\psi\rangle \in{\cal H}_1\otimes{\cal H}_1$ be a normalized
absolute ground state for $H'$ with $S'|\psi\rangle =m|\psi\rangle$.
We want to show that $m=0$. We may write $|\psi\rangle
=\sum_n|\psi_n\rangle$, where $S\otimes
I|\psi_n\rangle=n|\psi_n\rangle$ and $I\otimes
S|\psi_n\rangle=(n-m)|\psi_n\rangle$.

We may consider any state $|\phi\rangle \in{\cal H}_1\otimes{\cal
    H}_1$ as an operator from the Hilbert space ${\cal H}_1$ to itself,
in the following way. We introduce a basis $|X)$ in ${\cal H}_1$
indexed by subsets $X\subseteq \Lambda_L$, labeling the state with
all spins up at the sites in $X$ and down elsewhere. We shall refer to
this representation as the standard basis.  Then $|\phi\rangle$ may be
represented as a function associating a complex number $\phi(X,Y)$ to
any pair of subsets $X,Y\subseteq\Lambda_L$, namely $\phi(X,Y)$ is
given by the inner product of $|X)\otimes |Y)$ with $|\phi\rangle$.
Hence $|\phi\rangle$ may be identified with the operator
$\widehat\phi$ defined by the matrix elements $(X|\widehat\phi |Y)= \phi(X,Y)$.

If $A$ is an operator on ${\cal H}_1$ then $\widehat{A\otimes
    I|\psi\rangle} =A\widehat \psi$ and $\widehat{I\otimes
    A|\psi\rangle}=\widehat\psi A^T$ where $A^T$ is the {\it transposed}
operator represented by the matrix $A^T(X,Y)=A(Y,X)$ in the standard
basis. (Note that transposition is not a canonical operation, but
depends on the basis in which it is defined.)

The operator $S$ is represented by a real symmetric matrix in the
standard basis.  Thus in the above representation $\widehat\psi_n$
maps the subspace where $S=(n-m)$ to the subspace where $S=n$ and
vanishes on the orthogonal complement.  Hence we see that
$\widehat\psi^\dagger\widehat\psi=\sum_n
\widehat\psi_n^\dagger\widehat\psi_n$ and
likewise
$\widehat\psi\widehat\psi^\dagger=\sum_n\widehat\psi_n\widehat\psi_n^\dagger$.
It
follows from this that
$S\widehat\psi^\dagger\widehat\psi=\widehat\psi^\dagger\widehat\psi S$ and
$S\widehat\psi\widehat\psi^\dagger=\widehat\psi\widehat\psi^\dagger S$.  Hence
\begin{equation}\label{eq:S'=0}
S(\widehat\psi^\dagger\widehat\psi)^{1/2}=(\widehat\psi^\dagger\widehat\psi)^{1/2}
S,\quad
S(\widehat\psi\widehat\psi^\dagger)^{1/2}=(\widehat\psi\widehat\psi^\dagger)^{1/2} S.
\end{equation}

Let $|\psi_1\rangle,|\psi_2\rangle \in{\cal H}_1\otimes{\cal H}_1$
denote the states
such that $\widehat\psi_1=(\widehat\psi\widehat\psi^\dagger)^{1/2}$ and
$\widehat\psi_2=(\widehat\psi^\dagger\widehat\psi)^{1/2}$. Then
$|\psi_1\rangle$ and
$|\psi_2\rangle$ are normalized
(since $\langle \psi_1|\psi_1\rangle=\Tr(\widehat\psi_1^\dagger\widehat\psi_1)=
\Tr(\widehat\psi\widehat\psi^\dagger)=\langle\psi|\psi\rangle$)
and (\ref{eq:S'=0}) implies that $S'|\psi_1\rangle=0$ and $S'|\psi_2\rangle=0$.

We shall prove that
\begin{equation}\label{eq:reflectionpositivity}
     \langle \psi|H'|\psi\rangle \geq \frac12 \langle \psi_1|H'|\psi_1\rangle
+\frac{1}{2}\langle \psi_2|H'|\psi_2\rangle.
\end{equation}
Since $|\psi\rangle$ is an absolute ground state we see that
$|\psi_1\rangle$ and
$|\psi_2\rangle$ are also absolute ground states. Since they both have $S'=0$
and the ground state with this property is unique we conclude that
$|\psi_1\rangle=|\psi_2\rangle$, i.e.,
$\widehat\psi^\dagger\widehat\psi=\widehat\psi\widehat\psi^\dagger$. Then since
we are using a representation in which the matrix for $S$ is real and
symmetric we have
\begin{eqnarray*}
m&=&\langle\psi|S'|\psi\rangle=\langle\psi|(S\otimes I-I\otimes
S)|\psi\rangle\\
&=&\Tr(\widehat\psi\widehat\psi^\dagger S)-\Tr(\widehat\psi^\dagger\widehat\psi
S) =0.
\end{eqnarray*}

It remains to show the reflection positivity
(\ref{eq:reflectionpositivity}).  We may rewrite
\begin{multline}
\langle\psi|H'|\psi\rangle=\Tr(\widehat\psi\widehat\psi^\dagger H_L)+\Tr
(\widehat\psi^\dagger\widehat\psi
H_L)\\ -\half \sum_{\vecx\in M_L}
\left(\Tr(\widehat\psi^\dagger S^+_{\vecx}\widehat\psi
      S^-_\vecx) +\Tr(\widehat\psi^\dagger S^-_\vecx\widehat\psi
S^+_{\vecx})\right).
\end{multline}
Here $M_L$ denotes the set of sites in $\Lambda_L$ that connect to
a bond in $M$, i.e., sites in $\Lambda_L$ that are nearest neighbor to
a site in $\Lambda_R$.  We have used that the operators $S^\pm_{\vecx}$
are represented by real matrices in the standard basis.

The inequality (\ref{eq:reflectionpositivity}) now follows from the
inequality
\begin{eqnarray*}
\Tr(\widehat\psi^\dagger A\widehat\psi A^\dagger)&\!\leq&\!
\left[\Tr\left((\widehat\psi\widehat\psi^\dagger)^{1/2}
A(\widehat\psi\widehat\psi^\dagger)^{1/2}A^\dagger\right)\right]^{1/2}\\&&
\times
\left[\Tr\left((\widehat\psi^\dagger\widehat\psi)^{1/2}
A(\widehat\psi^\dagger\widehat\psi)^{1/2}A^\dagger\right)\right]^{1/2},
\end{eqnarray*}
which holds for any operator $A$.  This inequality is a simple
application of the Cauchy-Schwarz inequality if one uses polar
decomposition, i.e., the existence of a partial isometry $U$ such that
$$
\widehat\psi=U(\widehat\psi^\dagger\widehat\psi)^{1/2} \hbox{ and
}(\widehat\psi\widehat\psi^\dagger)^{1/2}=
U(\widehat\psi^\dagger\widehat\psi)^{1/2}U^\dagger.
$$

At positive temperature we may consider the partition function for $H$
restricted to the subspaces with fixed value of $\sum_\vecx
S^3_\vecx$.  We define
$$
Z(m)=\Tr \p_m\exp(-\beta H),
$$
where $\p_m$ is the projection onto the eigenspace of $\sum_\vecx S^3_\vecx$
corresponding to the eigenvalue $m$.
We claim that the partition function is maximal at half-filling, i.e.,
\begin{equation}\label{eq:maxpart}
      Z(m)\leq Z(0).
\end{equation}
To prove this we shall again use reflection positivity.

We first note that the unitary change which mapped $H$ to $H'$ will take
$\p_m$ into the operator
$$
\p'_m=\sum_n P_n\otimes P_{n-m},
$$
where $P_n$ is the projection operator in ${\cal H}_1$ projecting onto the
eigenspace of $S$ with eigenvalue $n$.
Observe that $P_m$ is a real matrix in the standard basis.

Using the Trotter product formula we may now write
\begin{multline}\nonumber
Z(m)=\lim_{k\to\infty}\Tr\p'_m\times \\ \left(e^{-\frac{\beta}{k}
H_L\otimes I}e^{-\frac{\beta}{k}I\otimes H_L}e^{\frac{\beta}{2k}\sum_{M}
S^+_\vecx S^+_\vecy} e^{\frac{\beta}{2k}\sum_{M}S^-_\vecx S^-_\vecy}\right)^k.
\end{multline}
If we use that
$$
e^{\frac{\beta}{2k}\sum_{\langle \vecx\vecy\rangle\in
M}S^\pm_\vecx S^\pm_\vecy}=\prod_{\langle
\vecx
\vecy\rangle \in
M}\left(1+\frac{\beta}{2k}S^\pm_\vecx S^\pm_\vecy\right)
$$
we see that the trace above may be written as sums of terms of the form
$A_nA_{n-m}$, where
$$
A_n={\rm Tr }_{{\cal H}_1}\, \left(P_n e^{-\beta H_L/k}T_1e^{-\beta
H_L/k}T_2\cdots
e^{-\beta H_L/k} T_k\right),
$$
and each of the operators $T_1,T_2\ldots$ is a monomial
in the variables $(\beta/2k)^{1/2}S_\vecx^\pm$, $\vecx\in M_L$.

Since $A_n$ is real for all $n$ we see that $A_nA_{n-m}\leq
A_n^2/2+A_{n-m}^2/2$. If we insert this above and simply undo the
calculation we arrive at (\ref{eq:maxpart}).

\section{Reflection positivity contour bound} \label{app:pathRP}

In this appendix we derive Lemma~\ref{lem:rp}, using reflection
positivity arguments.

The measures $\upsilon_z(d\omega)$ are reflection positive in the
following sense.  Draw a hyperplane, either vertically (through bonds) or
horizontally, that divides $\Gamma=\Lambda \times [0,\beta] $ into two
congruent parts $\Gamma_L\cup\Gamma_R$.  For any configuration
$\omega$, let $\bar \omega$ be its natural reflection through the hyperplane
(reversing its direction), and for any function $h$ on the space of
configurations let $\widetilde h(\omega) = h(\bar \omega)$.  Then, for
any such complex valued function that depends only on the restriction
of $\omega$ to $\Gamma_L$,
\begin{equation}  \label{h}
\int_{\Aa^{\rm o}}  h(\omega)  \widetilde h( \omega)^*\,
e^{-\lambda|\omega|}\,  \upsilon_z(d \omega) \,
   \geq  \,  0 \, .
\end{equation}
This can be seen by noting that once the behavior of $\omega$
on the hyperplane is fixed, the distribution of the  left and right 
sides (or top and bottom)
are conditionally independent, and are mirror images of
each other.

Reflection positivity leads to what is known as the {\em chessboard
   inequality} \cite{FS,FL}.  In essence, it is a multiply reflected
generalization of the Schwarz inequality, which allows us to obtain
bounds on the expectation value of a product of local variables in
terms of thermodynamic quantities.

The function whose average we need to estimate is $\chi_D(\omega)$ --
the indicator function which is $1$ if the curves in $\omega$ avoid a
specified set $D\subset \Gamma$ and $0$ otherwise.  One may start by
partitioning the imaginary time interval $[0,\beta]$, and
correspondingly the ``space-time'' $\Gamma$, into equal short segments
whose reflections tile $\Gamma$.  For any subset $D\subset \Gamma$
that is a union of elements of the finite partition of $\Gamma$ the
strategy, which is explained in detail in \cite{FL}, yields
\begin{multline}
\frac {\int _{\Aa^{\rm o}} \upsilon_{z} (d
\omega) \chi_D(\omega) e^{-\lambda |\omega|}}{\int _{\Aa^{\rm o}}
\upsilon_{z}(d \omega)
e^{-\lambda|\omega|}} \\ \label{chbound}  \le   \left(
\frac {\int _{\Aa^{\rm o}} \upsilon_{z} (d
\omega) \chi_\Gamma (\omega) e^{-\lambda |\omega|}}{\int _{\Aa^{\rm o}}
\upsilon_{z}(d \omega)
e^{-\lambda|\omega|}} \right)^{ |D|/|\Gamma|}  =  e^{ f |D|}  ,
\end{multline}
where $|D|$ is the  total length of $D$.
By refining the partition, and  applying elementary continuity arguments
(the dominated convergence theorem), we conclude that (\ref{chbound})
extends to all sets $D$ which are finite unions of closed intervals.
This proves  (\ref{claim2}).

To prove the second statement in Lemma~\ref{lem:rp}
we note that reflection positivity holds also for the restriction
of the measure to $\omega$'s with $0$ winding number.  I.e.,
for $h$ as in (\ref{h}),
$$
\int_{\Aa^{\rm o}}  h(\omega)  \widetilde h( \omega)^*\,   I[\nu(\omega)=0]\,
e^{-\lambda|\omega|}\,  \upsilon_z(d \omega) \
   \geq  \  0 \, .
$$
If the hyperplane dividing
$\Gamma$ into $\Gamma_L\cup\Gamma_R$ is horizontal, this is clear,
since fixing $\omega$ on this hyperplane fixes the winding number. If it is
vertical, however, we note that $I[\nu(\omega)=0]$ can be written as
$\sum_k I_k(\omega)\widetilde I_k(\omega)$, where $k$ runs from $0$ to
$|\Lambda|/2$, and $I_k$ is the indicator function for $\omega$
restricted to $\Gamma_L$ having winding number $k$. (Note that
$\omega$ restricted to $\Gamma_L$ may have sources and sinks on the
boundary, and when counting the winding number we also have to
consider the curves resulting from these.)  The second claim made in
Lemma~\ref{lem:rp}  follows by proceeding as in (\ref{chbound}), but with
the added restriction to $0$ winding number.

\end{document}